\documentclass[sigconf,screen]{acmart}
\AtBeginDocument{%
  }
\usepackage{xspace}
\usepackage{xcolor}
\usepackage{hyperref}
\usepackage{booktabs}
\usepackage{graphicx}
\usepackage{listings}
\usepackage[T1]{fontenc}
\usepackage{multirow}
\usepackage{float}
\usepackage{amsmath}
\usepackage{array}
\usepackage{longtable}
\usepackage{lipsum}
\usepackage{makecell}
\usepackage{xurl}
\acmBooktitle{Companion Proceedings of the 33rd ACM Symposium on the Foundations of Software Engineering (FSE '25), June 23--27, 2025, Trondheim, Norway}
\acmConference[FSE '25 Companion]{Companion Proceedings of the 33rd ACM Symposium on the Foundations of Software Engineering}{June 23--27, 2025}{Trondheim, Norway}

\newcommand{\longname}{Syntactic Prevalence Analyzer\xspace}
\newcommand{\name}{\textsc{SPA}\xspace}

\author{Jae Yong Lee}
\affiliation{%
  \institution{KAIST}
  \city{Daejeon}
  \country{South Korea}
}
\email{jaeyonglee0205@kaist.ac.kr}

\author{Sungmin Kang}
\affiliation{%
  \institution{KAIST}
  \city{Daejeon}
  \country{South Korea}
}
\email{sungmin.kang@kaist.ac.kr}

\author{Shin Yoo}
\affiliation{%
  \institution{KAIST}
  \city{Daejeon}
  \country{South Korea}
}
\email{shin.yoo@kaist.ac.kr}
\begin{document}

\title{Predictive Prompt Analysis} 

\begin{abstract}

Large Language Models (LLMs) are machine learning models that have seen widespread adoption due to their capability of handling previously difficult tasks. LLMs, due to their training, are sensitive to how exactly a question is presented, also known as prompting. However, prompting well is challenging, as it has been difficult to uncover principles behind prompting -- generally, trial-and-error is the most common way of improving prompts, despite its significant computational cost.
In this context, we argue it would be useful to perform `predictive prompt analysis', in which an automated technique would perform a quick analysis of a prompt and predict how the LLM would react to it, relative to a goal provided by the user. As a demonstration of the concept, we present \longname (\name), a predictive prompt analysis approach based on sparse autoencoders (SAEs). \name accurately predicted how often an LLM would generate target syntactic structures during code synthesis, with up to 0.994 Pearson correlation between the predicted and actual prevalence of the target structure. At the same time, \name requires only 0.4\% of the time it takes to run the LLM on a benchmark. As LLMs are increasingly used during and integrated into modern software development, our proposed predictive prompt analysis concept has the potential to significantly ease the use of LLMs for both practitioners and researchers.

\end{abstract}



\maketitle
\section{Introduction}
\label{sec:introduction}

Large Language Models (LLMs) are statistical models that predict the likelihood of the next token given preceding context, which have a large number of parameters and are trained on large corpus. An interesting characteristic of these models is that they show emergent task-solving capabilities when scaled ~\cite{wei2022emergentabilitieslargelanguage}, which has led to their widespread use in software engineering tasks ~\cite{10.1145/3695988}. In the most common use case of LLMs, one will describe the task in natural language, which is known as prompting the LLM.

Early experiments on LLMs demonstrated that the way one prompts an LLM has a significant influence on performance~\cite{10.1145/3560815}. 
However, it is often difficult to know which prompts will perform well in practice. As a result, prompt construction often involves significant trial-and-error~\cite{dang2022promptopportunitieschallengeszero} or prompt optimization based on ground-truth answers~\cite{pryzant2023automaticpromptoptimizationgradient}, both of which require a substantial level of human intervention and computational resources.

\begin{figure}[t!]
    \centering
    \includegraphics[width=1\linewidth]{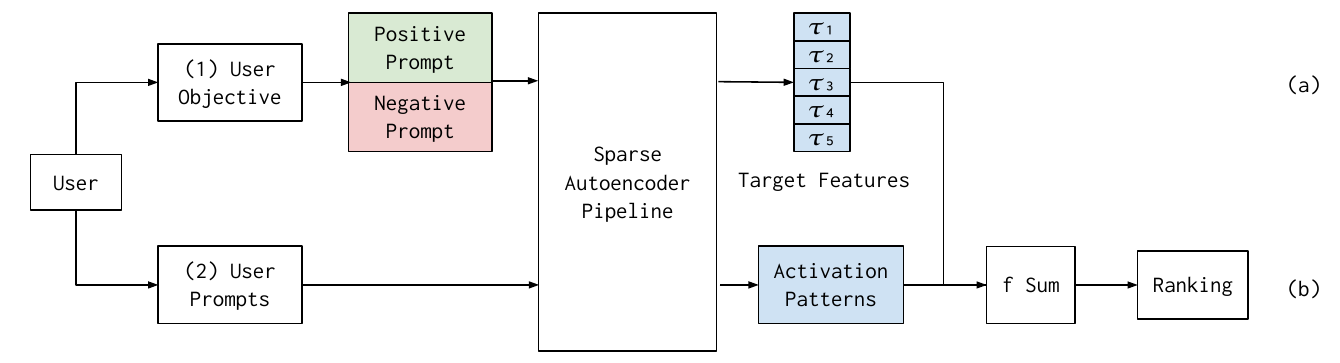}
    \caption{An Overview of \name, a Prototype Implementing Predictive Prompt Analysis. Upper Half (a): Extracting Target Features. Lower Half (b): Calculating the Ranking.}
    \Description{overview}
    \label{fig:overview}
\end{figure}

In response, we argue for \emph{predictive prompt analysis} -- quickly predicting the effect of a prompt before running it on a benchmark, without the need for user-side training or significant user input. By accurately forecasting the effect of prompts in a computationally inexpensive manner, developers could quickly and cheaply iterate on prompt designs that meet their goals. Such analyses would also grow in importance as state-of-the-art LLMs continue to scale, making their computational costs increasingly burdensome.

As a demonstration that making such a prediction cheaply and without user-side training is possible, we showcase \longname (\name), a preliminary predictive prompt analysis tool. We design \name around a simple scenario: the user seeks to have the LLM generate a target syntactic structure, such as a try-except clause, during code synthesis, but there are multiple ways to `ask' for such a structure; how effective would each prompt be in generating the target structure? \name predicts an answer to this question, orders of magnitude more quickly than running the prompt on a code synthesis benchmark. \name uses Sparse Autoencoders (SAEs), which are models that `cluster' the internal activation patterns of LLMs; 
specific clusters or `features' of SAEs can often be mapped to recognizable concepts~\cite{bricken2023monosemanticity}. \name first identifies the SAE features related to the user request, then based on these identified features predicts the relative incidence of the syntactic structures, ultimately allowing \name to predict how well each prompt would meet the goal.
Experiments demonstrate that \name shows strong predictive performance -- the actual incidence of the target syntactic structure for each prompt closely followed the predictions of \name, with a Pearson correlation value of up to 0.994; meanwhile, the time it took to run the predictive analysis was only 0.4\% of the total time to run the LLM on the full benchmark, and 18.7\% of a baseline that generated 10 code samples per instruction, demonstrating the significant computational efficiency of our approach.
Despite our strong early findings, there are two important limitations to \name. First, in this preliminary work, the behavior we predict, syntactic structure prevalence, is fairly artificial. Second, our approach currently only works on open-weight LLMs for which there is a trained SAE. Nonetheless, we have optimistic initial results suggesting that improved predictive prompt analysis tools could overcome these limitations of \name; such tools would ease rapid and computationally efficient prompt engineering.
In summary, we (i) propose the concept of predictive prompt analysis; (ii) describe the prototype tool \name implementing it; (iii) provide empirical results demonstrating the strong performance of \name; and (iv) describe the future directions that are promising.

\section{Related Work}

\paragraph{Prompt Engineering} Designing and refining how to ask the LLM to do the bidding of the user, without modifying any internal parameters, is known as prompt engineering. As LLMs are computationally expensive, the general practice is to optimize the prompt to achieve the most from a single LLM query ~\cite{sahoo2024systematicsurveypromptengineering}. Related techniques include Chain-of-Thought Prompting~\cite{10.5555/3600270.3602070}, few-shot prompting~\cite{brown2020language}, and Promptbreeder~\cite{10.5555/3692070.3692611}.
However, despite extensive research efforts, finding good prompts is mostly done via trial and error, due to the absence of explicit design principles that would help users construct an effective prompt. Therefore, the search for effective prompts is both challenging and expensive, often necessitating multiple LLM inference runs which are becoming more expensive as the models continue to scale~\cite{zhou2024surveyefficientinferencelarge}. Predictive prompt analysis, by reducing the computational time and developer waits currently involved in prompt optimization, thus has the potential to significantly ease the prompt engineering process.

\paragraph{Sparse Autoencoders}
One promising attempt at interpreting LLMs is the Sparse Autoencoder (SAE). SAEs decompose or `cluster' the internal activation of LLMs into \textit{features}, which tend to be easier to interpret than the neuron activation patterns~\cite{bricken2023monosemanticity}. In particular, SAEs take as input the residual stream values of an LLM at a particular token, and generate a sparse encoding where only a few (sparse) features are activated. Prior work shows that these SAE features can often be mapped to identifiable concepts, and thus be used for interpreting LLM behavior~\cite{bricken2023monosemanticity}. While research regarding SAEs is active~\cite{bricken2023monosemanticity, gao2024scalingevaluatingsparseautoencoders, huben2024sparse, templeton2024scaling}, they have not yet been used to analyze prompts as we propose. As prompts are the usual way developers interact with LLMs, predictive prompt analysis provides a unique vantage point to help developers using SAEs. While predictive prompt analysis is not restricted to the use of SAEs, we use them in \name as SAEs are a general way of capturing LLM activation patterns without the need for additional training by the user, and thus SAE features may be predictive of LLM behavior in turn.

\section{Methodology}
\label{sec:methodology}


\begin{table}[h]
  \centering
  \begin{tabular}{c|l}
      \toprule
      ID & Instructions \\
      \midrule
      1 & None \\
      2 & It might be helpful to add an exception handler. \\
      3 & Write an exception handler. \\
      4 & You need to write an exception handler. \\
      5 & Please, with all my heart, include an exception handler. \\
      \bottomrule
  \end{tabular}
  \caption{Exception Handler Instructions}
  \label{tab:instruction}
\end{table}

As described earlier, consider the scenario where a user wants a syntactic structure to be generated during LLM code synthesis, such that exception handlers are generated to prevent potential errors. The user is struggling between the options in Table~\ref{tab:instruction}, as it is unclear \textit{a priori} how consistently each instruction will lead to the LLM generating exception handling. Previously, the only way of answering this was to generate code based on these instructions hundreds or thousands of times. For example, even OpenAI suggests generating thousands of results to decide which prompt is better~\cite{openai_prompt_engineering}, which is both computationally and financially expensive.

To mitigate this substantial cost, we propose \name, a predictive prompt analysis technique which will predict which instruction is best. \name takes two inputs: (1) the description of a target syntactic structure in natural language (e.g. ``a try-except clause'') and (2) multiple prompts to compare, towards the goal of predicting which prompt will most consistently generate the target structure (shown in Tab.~\ref{tab:instruction}). With these inputs, \name goes through two phases: (a) \textit{Extracting the Target Features} and (b) \textit{Calculating the Ranking}, as illustrated in Fig.~\ref{fig:overview}, each described in detail next.

\subsection{Extracting Target Features}
\label{subsec:target_features}
\begin{figure}[h]
    \centering

    \framebox{\includegraphics[width=0.75\linewidth]{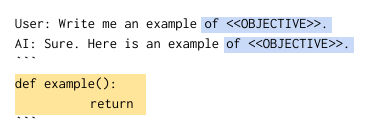}}
    \caption{Template of positive prompt}
    \Description{template}
    \label{fig:positive_prompt}
\end{figure}

To precisely extract the relevant SAE features, we employ `positive' and `negative' prompts. 
Figure~\ref{fig:positive_prompt} shows the template of a positive prompt, while removing ``of <<OBJECTIVE>>'' would make a negative prompt. 
Both prompts are needed to distinguish relevant SAE features related to the provided objective from generic features. 
For instance, on the prompt in Figure~\ref{fig:positive_prompt}, generic SAE features that activate on "python code" or the word "write" are activated along with task-relevant features, limiting the effectiveness of the extraction process.
Hence, a negative prompt, which has the same generic structure without mention of the objective, helps excluding these generic features. 
We extract features that activate from the example code, highlighted in yellow, to find features that influence the LLM in response to the instruction.

Formally, 
we rank features by the difference in activation strength between prompts, $d^\tau$:
\begin{equation}
d^{\tau} = \sum_{x \in T_{\text{pos}}^{\tau}} A_{x}^{\tau} - \sum_{y \in T_{\text{neg}}^{\tau}} A_{y}^{\tau}
\end{equation}
where $T_{\text{pos}}^{\tau}$ and $T_{\text{neg}}^{\tau}$ are the set of activated tokens on a feature $\tau$ for positive and negative prompts, respectively, and $A_{x}^{\tau}$ is the activation value of feature $\tau$ on the token $x$.
We employ the ranking process on the basis that features with highly differing activation values are the most relevant with the provided context~\cite{templeton2024scaling}; 
Ranking by $d^\tau$, \name selects the top $k$ features to acquire an SAE feature set $TF$, which are ideally related to the target syntactic structure.

\subsection{Calculating the Ranking}

Once relevant features have been extracted, they are used to predict the effectiveness of the instructions. For each instruction, a set of prompts $S$ is formed for code synthesis, in which the instruction is combined with randomly sampled problems from a code synthesis dataset.
While \name could operate based on a single example code synthesis problem in principle, we aggregate the results of analysis over multiple problems as in Eq.~\ref{eq:value_equation} to reduce noise.

The activation patterns of the LLM over prompts $S$ are then processed by the SAE, yielding the final prediction. First, we define the normalized activation frequency as $f^\tau = \frac{t_{\text{activated}}}{t_{\text{total}}}$, where $t_{\text{activated}}$ is the number of tokens for which the target feature $\tau$ is activated and $t_{\text{total}}$ is the total number of tokens. Using this, the prediction $P$ for each instruction is defined as:

\begin{equation} \label{eq:value_equation}
P = \sum_{\tau \in TF} \sum_{p \in S} f^\tau_p
\end{equation}

\noindent where $TF$ is the target feature set from Section~\ref{subsec:target_features}, $S$ is the sample set of prompts, and $f_p$ is the normalized frequencies calculated from the encoded prompt $p$. Ultimately, instructions are ranked in descending order of $P$ and presented to the user, as we expect that instructions causing the relevant features to activate will more likely generate the target syntactic structures.

\section{Experimental Design}
\label{sec:design}

\subsection{Research Questions}

This study answers the following research questions:
\begin{itemize}
    \item \emph{RQ1 (Efficacy)}: How effective is \name in predicting the effect of prompts?
    \item \emph{RQ2 (Efficiency)}: What is computational overhead of \name in predicting the effect of prompts? 
\end{itemize}

\subsection{Experimental Setup}

Experiments were conducted using AMD EPYC 9124 16-Core Processor X86\_64 CPUs, and four NVIDIA GeForce RTX 4090 GPUs.
We used the \textsc{Google Gemma-2-2b-it}~\cite{gemmateam2024gemma2improvingopen} model with \textsc{Gemma-2-2b SAE} ~\cite{lieberum-etal-2024-gemma} that has 16384 trained features. SAEs are trained on specific layers of LLMs to analyze activation patterns at that layer; we used SAEs trained on layer 1 (first layer), 9, 16, and 25 (last layer) to further analyze the impact of layer depth on the prediction performance of our technique.

Three syntactic objectives are used: generating try-except clauses, comments, and print statements. For each objective, both instruction sets and 
fewshot prompts were manually curated through author consensus, simulating a prompt engineering process of involving instructions with varying tones.
 We experiment with six prompt sets, also listed in Tab.~\ref{tab:results}. \textit{Exception} and \textit{Print} are composed of five instructions varying degrees of authoritative tone (see Fig.~\ref{tab:instruction}), while \textit{Comment} is composed of eight instructions with three designed to suppress comment generation as the LLM tended to produce comments even when given no instructions. 
 For non-fewshot scenarios, an empty prompt was inserted to observe the inherent tendency of LLM generations. 
 \textit{Exception Fewshot} range from 0-shot to 4-shot, whereas \textit{Comment Fewshot} and \textit{Print Fewshot} are composed up to 3-shot due to memory constraints of our environment. 
 The instructions for \emph{Exception} can be found in Tab~\ref{tab:instruction}.

As a benchmark to evaluate our technique we used the sanitized version of Mostly Basic Python Problems (MBPP) which has 427 natural language specifications and corresponding code, 
a manually verified set of crowd-sourced Python programming problems~\cite{austin2021programsynthesislargelanguage}.
For \name, we sampled 10 random problems ($|S| = 10$) from MBPP and merged them with each instruction. As mentioned in Section~\ref{subsec:target_features}, 
five target features were extracted from the prompts ($|TF| = 5$). Pearson correlation for \name's predictions was computed against LLM inference outputs on the entire MBPP dataset, 
averaged over three runs, with the total number of times the target syntactic structure was generated overall tallied. 
A strong correlation would suggest that 
\name effectively captures the causal dependencies between prompt and LLM behavior, 
thus validating its reliability as a quantitative metric for the task in question.
We also compare against a partial inference baseline that involves sampling 10 inference outputs from the same 10 problems used by \name and counting target syntactic structures. 
For both the partial inference and \name predictions, the average was calculated on Fisher Z-transformed correlations over five attempts~\cite{7e2958c8-cf46-3edc-b197-57ee54882a19}.

\section{Results}
\label{sec:result}

\subsection{RQ1: Effectiveness of \textit{SPA}}
\label{subsec:correlation}

\begin{table}[t]
    \caption{Average Correlations Across Different Scenarios and Layers Compared to the Sampled Inference}
    \label{tab:results}
    \renewcommand{\arraystretch}{1.1}
    \setlength{\tabcolsep}{3pt}
    \begin{tabular}[width=0.7\linewidth]{lccccc}
    \toprule
    \multirow{2}{*}{\textbf{Scenario}} & \multirow{2}{*}{\textbf{\shortstack{Sampled \\ Inference}}} & \multicolumn{4}{c}{\textbf{Layer \#}} \\ 
    \cmidrule(lr){3-6}
     &  & \textbf{1} & \textbf{9} & \textbf{16} & \textbf{25} \\ 
    \midrule
    Exception          & 0.743 & 0.972 & 0.633 & 0.978 & 0.790 \\
    Exception Fewshot  & 0.744 & 0.956 & 0.952 & 0.959 & 0.954 \\
    Comment            & 0.743 & 0.687 & 0.887 & 0.751 & 0.860 \\
    Comment Fewshot    & 0.740 & 0.941 & 0.937 & 0.908 & 0.926 \\
    Print              & 0.740 & 0.960 & 0.971 & \textbf{0.994} & 0.913 \\
    Print Fewshot      & 0.725 & 0.953 & 0.957 & 0.923 & 0.957 \\
    \hline
    \emph{Average}            & 0.740 & 0.911 & 0.890 & \textbf{0.918} & 0.900 \\
    \bottomrule
    \end{tabular}
\end{table}

In Table~\ref{tab:results}, we observe that the partial inference baseline we compare against achieves an average correlation of 0.740. In contrast, over all layers, \name has a better average correlation with the occurrence of the target syntactic structure. Among the layers, layer 16 shows the most best correlation across all scenarios, also with the highest correlation: 0.994 for \textit{Print}. Based on these results, it appears to be more effective to utilize SAEs trained on layers positioned approximately $\frac{2}{3}$ of the way (layer 16) into the network, similar to previous work on SAEs that find layers between $\frac{1}{2}$ and $\frac{5}{6}$ of the way to be the most interpretable~\cite{templeton2024scaling, gao2024scalingevaluatingsparseautoencoders}.
\newline
\newline
\setlength{\fboxrule}{0.5mm}
\framebox[\linewidth][l]{
  \parbox{0.95\linewidth}{
    \textbf{Answer to RQ1}: \name achieves the highest average correlation of 0.918 using SAE trained with layer 16, outperforming the average correlation of the sampled inference output, 0.740.
  }
}
\newline

\subsection{RQ2: Efficiency of \textit{SPA}}
\begin{table}[t]
    \caption{Average Computation Time in Seconds}
    \label{tab:time_taken}
    \renewcommand{\arraystretch}{1.1} 
    \setlength{\tabcolsep}{1.3pt} 
    \begin{tabular}[width=\linewidth]{lcccc>{\bfseries}c}
        \toprule
        \multirow{2}{*}{\textbf{Scenario}} & 
        \multirow{2}{*}{\makecell{\textbf{Total} \\ \textbf{Inference}}} & 
        \multirow{2}{*}{\makecell{\textbf{Sampled} \\ \textbf{Inference}}} &
        \multicolumn{3}{c} {\textbf{\name}} \\
        \cmidrule(lr){4-6}
        & & & \textbf{Ext.} & \textbf{Pred.} & \textbf{Total} \\

        \midrule
        Exception          & 9428 & 195 & 0.212 & 38.1 & 38.3 \\ 
        Exception Fewshot  & 9409 & 239 & 0.218 & 44.6 & 44.8 \\ 
        Comment            & 14022 & 308 & 0.212 & 55.1 & 55.3  \\
        Comment Fewshot    & 9102 & 224 & 0.221 & 36.4 & 36.6 \\
        Print              & 8850 & 175 & 0.213 & 40.9 & 41.1 \\ 
        Print Fewshot      & 8541 & 209 & 0.214 & 35.3 & 35.5 \\ 
        \hline
        \emph{Average}            & 9892 & 225 & 0.215 & 41.7 & 42.0 \\
        \bottomrule
    \end{tabular}
\end{table}

Table~\ref{tab:time_taken} shows the time taken for prediction, in seconds, for different approaches of evaluating prompts. Notably, while the sampled inference requires an average of 225 seconds, \name required only 42.0 seconds, a decrease of 81.3\% despite showing better predictive performance, underscoring its efficiency and demonstrating the potential of predictive prompt analysis in reducing both the computational and financial overhead in prompt engineering. 
Furthermore, the average total inference time for 427 data points in MBPP was 9892 seconds, more than 235 times longer than \name. 
In terms of the time cost of each phase of \name, feature extraction took 0.215 seconds on average, while generating predictions for all instructions took 41.7 seconds on average.
\newline
\newline
\framebox[\linewidth][l]{
  \parbox{0.95\linewidth}{
    \textbf{Answer to RQ2}: \name is significantly more efficient than the sampled inference and total inference, showing a decrease of 81.3\% and 99.6\% of computation time, respectively.%
  }
}
\newline

\section{Qualitative Analysis}
\label{subsec:features}

In this section, we analyze the extracted features to investigate the effectiveness of the target feature extraction process described in Section~\ref{subsec:target_features}. As \name does not internally assess the quality of features other than their activation values, manual inspection of the features is helpful in understanding the process. For analysis we use Neuronpedia, a platform that allows visualization of SAE data \footnote{https://www.neuronpedia.org/}, to find the maximum activating examples from training data of \textsc{Gemma-2} composed of web documents, code, and math problems, on our extracted target features. 

Inspecting the results from layer 16 and from the `Exception' scenario, for which the correlation is high, \name identified SAE feature \#2423 as relevant, activates on the token "throw" in snippets such as \texttt{error.response = response; \textbackslash n throw}. This feature is clearly related with exception-related behavior, suggesting that it accurately captures specific characteristics of the objective. Similarly, for the \verb|python print statement| objective, \name identified SAE feature \#4961, which activates on snippets like \texttt{System.out.println}. This inspection shows that \name can identify target-relevant features, which in turn help predict the prevalence of syntactic structures.

\section{Limitations and Future Directions}
\label{sec:limitation}

Our current approach is limited to open-weight LLMs, such as \textsc{Gemma}, which allow the extraction of internal activations to train and use SAEs. However, state-of-the-art LLMs such as GPT and Claude do not allow access to the weights of the model, much less SAEs that use LLM activation, restricting the scope of \name. 
Although surmounting this may seem a significant challenge, we hope to experiment with the transferability of predictive prompt analysis from one LLM to another, similarly to the phenomenon of adversarial example transfer in image classifiers~\cite{papernot2016transferabilitymachinelearningphenomena}. Our preliminary experiments on similar LLMs, specifically between Gemma-2b and Gemma-2-2b, have shown promise, raising the possibility that we could perform predictive analysis on open-weight models and apply the results to closed-weight models.

Another limitation of \name is that it focuses on predicting the occurrence of syntactic structures. Practitioners and researchers would likely be more interested in comparatively abstract or semantic properties, such as general syntactic well-formedness or correctness. Thus, predictive prompt analysis tools that work on more semantic behavior would ideally be devised. We again have early preliminary results in this regard -- in our experiments with try-except clause prediction, we found features related to the abstract concept of `trying again', which would cause the LLM to suggest multiple solutions instead of a single solution. Being able to reliably identify such features would allow developers greater control over their prompt construction process, significantly improving the utility of predictive prompt analysis.

\section{Conclusion}
\label{sec:conclusion}

This work introduces the concept of predictive prompt analysis, in which a quick analysis is performed to predict how an LLM will behave in response to a prompt. Computationally efficient predictive prompt analysis techniques would ease development of LLM-based applications by accelerating the trial-and-error process of prompt engineering. We demonstrate that predictive prompt analysis is feasible through our prototype tool \name, which predicts prompt effect on generating syntactic structures by first identifying relevant features from SAEs, then using those features to analyze how the LLM will respond to prompts, all without the need for training. \name shows strong predictive performance while being computationally efficient: initial results across six selected scenarios show that the predictions made by \name of the relative prevalence of target syntactic structures achieve the highest average correlation of 0.918 with actual inference outputs, with up to a 99.6\% reduction in computation time. While \name, as a tool, has limitations that make it difficult to immediately apply to practical tasks, better predictive prompt analysis tools could potentially overcome them; in turn, predictive prompt analysis would become viable in practical settings and ease development of LLM-based software.

\section{Data Availability}
The code and data used in the paper is available from this \href{https://zenodo.org/records/14666485?token=eyJhbGciOiJIUzUxMiJ9.eyJpZCI6IjRjMmI1NmFiLWY5NzYtNGY5My1iZmVmLTI5N2M5YWUzMzliZSIsImRhdGEiOnt9LCJyYW5kb20iOiI4NjBlNTdlNjZkNjcwMWJmMzUwMDlmMGUwZTAxZTU5NSJ9.w7E8YaphtSrc5FygISjANwpG93hpQzvLUTU88bmG0vuChzWECk5WOeX3iHeS-uH1Dp7DLGoW4g9bahhtCqZaLw}{link}.

\newpage

\bibliographystyle{ACM-Reference-Format}

\bibliography{sample-base}

\end{document}